# A Novel Self-Recognition Method for Autonomic Grid Networks Case Study: Advisor Labor Law Software Application

Mehdi Bahrami[1], Peyman arebi[2], Hosseyn Bakhshizadeh[3], Hamed Barangi[3]
[1]Department of Computer Eng., Islamic Azad University, Booshehr Branch, Iran
E-mail: Bahrami@LianPro.com
[2]The Holy Prophet Higher Education Complex
[3]Department of Computer Eng, Islamic Azad University, Siahkal Branch, Iran
doi : 10.4156/aiss.vol3.issue5.30

## Abstract

*Recently, Grid Computing Systems have provided wide integrated use of resources. Grid computing systems provide the ability to share, select and aggregate distributed resources as computers, storage systems or other devices in an integrated way. Grid computing systems have solved many problems in science, engineering and commerce fields. In this paper we introduce a self-recognition algorithm for grid network and introduced this algorithm to have exclusive management control on the autonomic grid networks. This algorithm is base on binomial heap to allocate and recognition any node in the grid. We try to using this algorithm in advisor labor law software application as case study and shown in this application how to use this method for any advisor application on the network. By this implementation model shown this method can get better answer to any question as a best labor law advisor.*

## 1. Introduction

The physical structure of "Grid Computing System" is much similar to that of clusters, but the main difference is that instead of having dedicated servers for providing each service, services are provided by the user of the system itself [1].

In general "Grid Computing System" [2] acts as mediator between provider of the service and user of the service & for that provider of the services will receive some cash and user will have to pay for it and as mediator the Company acting as a mediator will get some commission from both provider and user of the services. A form of networking, Unlike conventional networks that focus on communication among devices, grid computing harnesses unused processing cycles of all computers in a network for solving problems too intensive for any stand-alone machine. It allows unused CPU capacity in all participating machines to be allocated to one application that is extremely computation intensive and programmed for parallel processing. Grid computing enables the virtualization of distributed computing [22] and data resources such as processing, network bandwidth and storage capacity to create a single system image, granting users and applications seamless access to vast IT capabilities. Just as an Internet user views a unified instance of content via the Web, a grid user essentially sees a single, large virtual computer.

Grid computing is an emerging computing model that treats all resources as a collection of manageable entities with common interfaces to such functionality as lifetime management, discoverable properties and accessibility via open protocols [19]. The Internet was developed to meet the need for a common communication medium between large, federally funded computing centers. These communication links led to resource and information sharing between these centers and eventually to provide access to them for additional users [15]. Ad hoc resource sharing 'procedures' among these original groups pointed the way toward standardization of the protocols needed to communicate between any administrative domains. The current grid technology can be viewed as an extension or application of this framework to create a more generic resource sharing context.

Grid computing offers a model for solving massive computational [20] problems by making use of the unused resources (CPU cycles and/or disk storage) of large numbers of disparate computers, often desktop computers, treated as a virtual cluster embedded in a distributed telecommunications infrastructure [21]. Grid computing focus on the ability to support computation across administrative domains sets it apart from traditional computer clusters or traditional distributed computing. Grid computing has the design goal of solving problems too big for any single supercomputer, whilst retaining the flexibility to work on multiple smaller problems. Thus Grid computing provides a multi-user environment. Its secondary aims are better exploitation of available computing power and catering for the intermittent demands of large computational exercises.

In this paper we try extended last published binomial heap [17] method for self-recognition in gird computing networks [16] and try to use this method in advisor labor law software application as case study.

Computing systems are expected to be effective. This means that serve a useful purpose when they are first introduced and continue to be useful as conditions change. Autonomic Computing [7], launched by IBM in 2001[18].

In 1998, Ian Foster and Carl Kesselman provided an initial definition in their book *The Grid: Blueprint for a New Computing Infrastructure* [11]: "A computational grid is a hardware and software infrastructure that provides dependable, consistent, pervasive, and inexpensive access to high-end computational capabilities."

An autonomic system is self-managing, meaning that it is self-protecting, self-configuration, self-healing and self-optimizing [2, 4].

The goal to virtualizes the resources on the grid and more uniformly handle heterogeneous systems will create new opportunities to better manage a larger, more distributed IT infrastructure. It will be easier to visualize capacity and utilization, making it easier for IT departments to control expenditures for computing resources over a larger organization [4].

The grid offers management of priorities among different projects. In the past, each project may have been responsible for its own IT resources and the associated expenses. Often these resources might be underutilized another project finds itself in trouble, needing more resources due to unexpected events. With the larger view a grid can offer, it becomes easier to control and manage such situations. Administrators can change any number of policies that affect how the different organizations might share or compete for resources.

Aggregating utilization data over a larger set of projects can enhance an organization's ability to project future upgrade needs. Autonomic computing can come into play here too. Various tools may be able to identify important trends throughout the grid, informing management of those that require attention [4].

It is assumed that an autonomic computing system is made up of a connected set of *autonomic elements*. Each element must include sensors and effectors [16].

Monitoring behavior through the sensors, comparing this with expectations, deciding what action, if any, is needed and then executing that action through effectors, creates a control loop (Figure 1) [7].

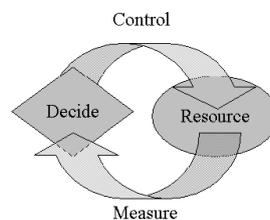

**Figure 1.** Control Loop [7]

## 2. Self-Recognition

We introduced self-recognition in autonomic grid networks require to a method for recognition any bottom of nodes in the grid networks, then we captive to many nodes in the network.

We can use many algorithms to reach this idea for this recognition but we want to introduce Binomial Heap for this and survey.

*1) Binomial Tree*

The binomial tree is the building block for the binomial heap. A binomial tree is an ordered tree – that is, a tree where the children of each node are ordered. Binomial trees are defined recursively, building up from single nodes. A single tree of degree k is constructed from two trees of degree k-1 by making the root of one tree the leftmost child of the root of the other tree.

*2) Binomial Heaps*

Binomial heaps were introduced in 1978 by Vuillemin [9]. Brown [10] studied their properties in detail [6].

A Binomial heap H is a set of binomial trees that satisfies the following binomial-heap properties.

    1. Each binomial tree in H obey the min-heap property: the key of a node is greater than or equal to the key of its parent. We say that each such tree is min-heap-ordered.

    2. For any non-negative integer k, there is at most one binomial tree in H whose root has K degree.

The first property tells us the root of a min-heap-ordered tree contains the smallest key in the tree.

The second property implies that an n-node binomial heap H consist of at log n+1 binomial trees.

The algorithms presented later work on a particular representation of a binomial heap. Within the heap, each node stores a pointer to its leftmost child (if any) and its rightmost sibling (if any). The heap itself is a linked list of the roots of its constituent trees, sorted by ascending number of children.

*3) Why Binomial Heaps?*

The Grid also involves dynamic, multi-institutional Virtual Organizations (VOs), where these new communities overlay classical organization structures, and these

Virtual organizations may be use as large or small, static or dynamic [12], then if this organization needs to be what node is exit and re-organization, then we need define, delete and update any node in this network by any node, because any node need to have some information about near node and collaborate when this node cant self-repairing [13] as dynamically, then this node can repair by top of node and auto repair.

If any node wants repair other node then need to recognition bottom level of node in the grid networks, for rich to this use Binomial Heap because this heaps can modify by pointer and have good worst-case time. [6]

*4) Binomial Heaps for Self-Recognition*

To self-recognition on the grid networks, any node must be having a table of bottom level on the grid; we define any node if it's adjacent mark as child of node.

Other adjacent, adjacent node it's mark as grandchild and so. But many nodes can child for many nodes and father for other nodes because we have a network not a tree.

As illustrated in Figure 2 we have a grid network, for allocate table recognitions for node 2 as Figure 3 is a more detailed representation of binomial heap H.

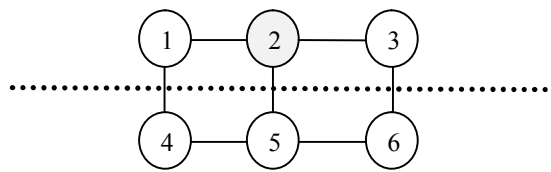

**Figure 2.** A Grid Network

In this figure it's only binomial heap for node 2 and we start at node 2, and allocate Head [H] to this node, at the next step recognition all adjacent as children for node 2. In this figure we shown No.1 and No.2 and No.3 nodes are children for No.2 node.

After this step we must recognition all adjacent, adjacent of node 2 or grandchildren for this node. Then we found node 4 for node 1 and node 5, node 6 it's a child for node 3 and node 5. Beside node 5 it's a child for node 4 and node 6.

In this allocation we have a note, that's node 2, its root then we must survey of the root and never draw any cycle to root because some cycle to root, then we can't find and recognition root of binomial heap.

*5) Operations on binomial heaps*

All operations on binomial heaps in Autonomic grid networks it's similar to normal operations on binomial heaps but we need a procedure to exploring nodes and insert any new nodes to binomial heap. For this

  *a) Create a new binomial heap*

To make an empty binomial heap, the MAKE-BINOMIAL-HEAP procedure simply allocates and return an object H, where head [H] =NIL. The running time is $\Theta(1)$ [6].

  *b) Insert node*

To insert a node in heap H, the INSERT (H, x) insert node x, whose key field has already been filled in, into heap H. The running time is O (lg n)[6].

  *c) Delete Node*

To delete a node from heap H, the DELETE (H, x) deletes node x from heap H. The running time is $\Theta$ (lgn)[6].

*6) Self-modify of Binomial Heaps*

It is even more important that this requirement be considered when an autonomic system can create new implicit versions of itself through self-modification. [3] Then we can have a binomial heap for every node. We have many solutions for update every node binomial heap; update every time is the best and have minimum running time.[8] Then we can update every node on the special time for gather new branch and delete every node in every branch.

## 3. Excusive Management

The goal to virtualizes the resources on the grid and more uniformly handle heterogeneous systems will create new opportunities to better manage a larger, more disperse IT infrastructure. It will be easier to visualize capacity and utilization, making it easier for IT departments to control expenditures for computing resources over a larger organization. [14]

Aggregating utilization data over a larger set of projects can enhance an organization's ability to project future upgrade needs. When maintenance is required, grid work can be rerouted to

Other machines without crippling the projects involved [8].

Autonomic computing can come into play here too. Various tools may be able to identify important trends throughout the grid, informing management of those that require attention [3, 4].

Any grid system has some management components. First, there is a component that keeps track of the resources available to the grid and which users are members of the grid. This information is used primarily to decide where grid jobs should be assigned.

Second, there are measurement components that determine both the capacities of the nodes on the grid and their current utilization rate at any given time. This information is used to schedule jobs in the grid. Such information is also used to determine the health of the grid, alerting personnel to problems such as outages, congestion, or over commitment. This information is also used to determine overall usage patterns and statistics, as well as to log and account for usage of grid resources.

Third, advanced grid management software can automatically manage many aspects of the grid. This is known as "autonomic computing," or "recovery oriented computing." This 16 Fundamentals of Grid Computing software would automatically recover from various kinds of grid failures and outages, finding

Alternative ways to get the workload processed. Then we try to solve any problem in this node by autonomic specification at the next step if this operation is fail other node by recognition this try to excusive management in top of level of this node.

We try simulating this algorithm for 100 grid nodes on the network, and get better result vs. non-binomial-heap algorithm, shown in figure 3.

## 4. Case STUDy: Advisor Labor Law Software Application

Advisor labor law [17] software application stands for "employment laws assistance for workers and small businesses". This software was developed to help employers and workers understand their rights and responsibilities under the federal employment laws administered. The Advisor labor law software application, Advisors mimic the interaction an individual might have with a representative by asking questions, providing information, and directing the individual to the appropriate resolution [17].

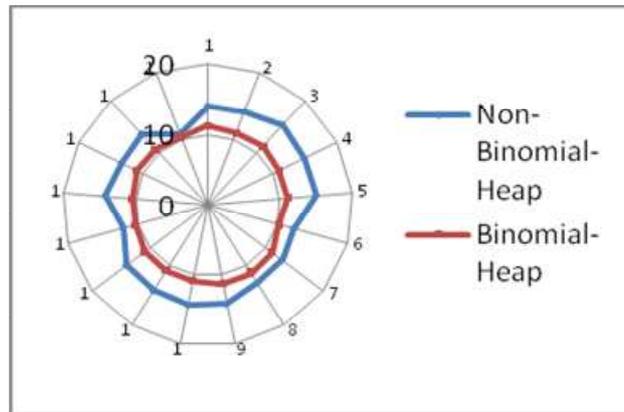

**Figure 3.** Result of 100 nodes in Grid network result by binomial-heap algorithm and none binomial-heap algorithm

Advisor labor law software application is one of a number of Parlian's Tools [1] developed to further dedication to provide clear, accurate and easy-to-access compliance assistance for employers and protect the wages, health benefits, retirement security, safety and health of Iranian's workforce.

This application is highly adept at counseling and representing management in all types of labor and employment matters, including:
- Overtime and other wage and hour claims;
- Discrimination and harassment issues;
- Wrongful dismissal claims;
- Non-compete, compensation and other employment agreements;
- Union collective bargaining agreements, unfair labor practices and grievances; and
- Other employment related concerns, such as leave law issues, disciplinary problems and employee theft.

In this application we need some information gather from other nodes and all advisor in any location should be use information from all other node but in this application, any node has a self-management on the network.

As illustrated in Figure 4, we used this method in advisor labor law software application implementation model. In this implementation model designed an autonomic grid management by exclusive and self-recognition for other nodes, after this if any node of advisor for better advise can't representation best answer then other top level nodes can recognition and try to present better answer.

In Figure 5, we try to using this method in C#.Net 3.5 by visit other child nodes, and then if Current State as current configuration is changed by child nodes, hence, new configuration that changed and saves in Current State and run this progress by this configuration. In this figure, you can find child nodes implementation by introduced binomial heaps.

---

[1]Software application tools developed in Advance Lab Intelligence Software, Lian Processor Co, Booshehr, Iran.

## 5. Related work

To the best of our knowledge, it's very little research contributions have been published toward applying autonomic computing principles in the context of self-recognition and excusive management in autonomic grid networks.

As proposal for related work we can offer use this method in other type of networks for recognition and management network as so as autonomic grid in this paper and we can transfer data with XMLR [5] method.

## 6. Conclusion

In the scope of the work presented here, we proposed an autonomic grid computing management algorithm by exclusive and self-recognition method for control of other nodes, in this paper we shown how can we use this algorithm in a software application as the related worked in the case study, we shown how can we used this method for any advisor software application on the networks and how can control and establish this method after this if any node can't repair then other top level nodes can recognition and try to repair this node in autonomic grid, and in case study, if any advisor can't present best answer, try to use this method for find better answer in top level advisor.

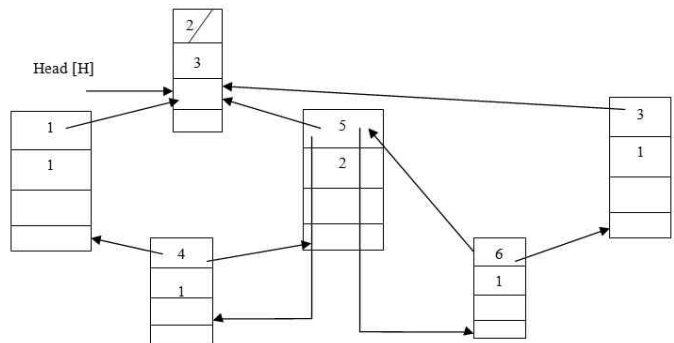

**Figure 3**. Binomial Heap definition for node 2 on the grid network

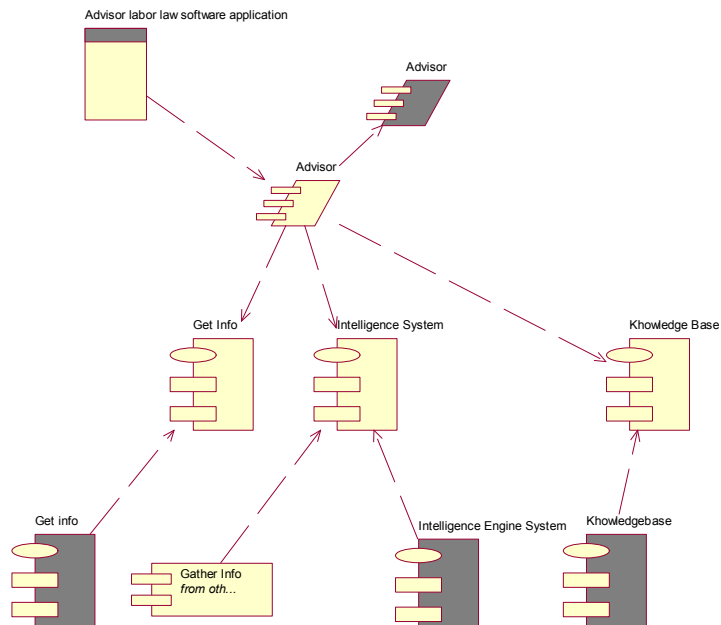

**Figure 4**. Implementation model of Advisor labor law software application

```csharp
private void Retrieve(object sender, EventArgs e, Stateobject CurrentState)
        {
            int         get_Child_Statement;

            if (get_Child_Statement > 0)
            {
                if (send_state_to_childs(CurrentState))

                    if (CurrentState != NewCurrentState)
                        Prgress(NewCurrentState);

                    else
                        Prgress(CurrentState);

            }
        }
```

**Figure 5**. Partial code in C# of Advisor labor law software application for run a rule and find better answer from other child nodes

**Mehdi Bahrami**

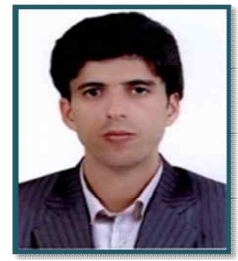

Mehdi Bahrami was born on July 13<sup>th</sup> in 1982. Mehdi received his Associate Level and B.S. degree in Software Engineering 2003 and 2007, respectively. The M.S. degree in Software Engineering from the Payam Noor University of Tehran has been accomplished in 2010.

Mehdi is Software Engineering lecturer university where he has served on the faculty since 2008. His areas of expertise are in Grid Computing Software Architecture, Software Architecture, Software Engineering, Compiler Design, and Automata Theory, Languages and Computation.

He has worked on the Software Engineering Methodology and Software Architecture in his Master and he hope can be continues his research in PhD level. He also investigates Intelligence Software Application Lab in Lian Processor Co. since 2007.

It worth mentioning that he is a member of technical program committee and reviewer in many international conferences. Mehdi has published many papers as first author and has taken the responsibility of over 60 final thesis and projects as an advisor in Associate level and contracts with universities and industry.